\documentclass{aa}
\usepackage[varg]{txfonts}
\usepackage[utf8]{inputenc}
\usepackage{graphicx}
\usepackage{array}
\usepackage{fp}
\usepackage{siunitx}
\usepackage{pdflscape}
\usepackage{numprint}
\usepackage{longtable}
\usepackage{multicol}
\usepackage{rotating}
\usepackage{microtype}

\begin{document}

\title{Derivation of parameters for 3748 FGK stars using H-band spectra from APOGEE Data Release 14}

\author{Pedro Sarmento \inst{1,2}
\and Elisa Delgado Mena \inst{1}
\and Bárbara Rojas-Ayala \inst{3}
\and Sergi Blanco-Cuaresma \inst{4}}

\institute{Instituto de Astrofísica e Ciências do Espaço, Universidade do Porto, CAUP, Rua das Estrelas, PT4150-762 Porto, Portugal e-mail:
pedro.sarmento@astro.up.pt
\and Departamento de Física e Astronomia, Faculdade de Ciências, Universidade do Porto, Portugal
\and Instituto de Alta Investigación, Universidad de Tarapacá, Casilla 7D, Arica, Chile
\and Harvard-Smithsonian Center for Astrophysics, 60 Garden Street, Cambridge, MA 02138, USA }

\abstract
{The Apache Point Observatory Galactic Evolution Experiment (APOGEE) has observed the H-band spectra of over 200 000 stars with $R\sim22 000$.}
{The main motivation for this work is to test an alternative method to the standard APOGEE pipeline (APOGEE Stellar Parameter and Chemical Abundances Pipeline, ASPCAP) to derive parameters in the Near-InfraRed (NIR) for FGK dwarfs.}
{\textit{iSpec} and \textit{Turbospectrum} are used to generate synthetic spectra matching APOGEE observations and to determine the parameters through $\chi^2$ minimization.}
{We present spectroscopic parameters ($T_\mathrm{eff}$, $[M/H]$, $\log g$, $v_{mic}$) for a sample of 3748 main-sequence and subgiant FGK stars, obtained from their APOGEE H-band spectra}
{We compare our output parameters with the ones obtained with ASPCAP for the same stellar spectra, and find that the values agree within the expected uncertainties. A comparison with the optical samples California Planet Survey, HARPS-GTO (High Accuracy Radial Velocity Planet Searcher - Guaranteed Time Observations), and PASTEL, is also available, and median differences below 10\,K for $T_\mathrm{eff}$ and 0.2\,dex for $[M/H]$ are found. Reasons for these differences are explored. The full H-band line-list, the line selection for the synthesis and the synthesized spectra are available for download, as well as the calculated parameters and their estimated uncertainties.}


\maketitle

\section{Introduction}

The analysis of light spectra from stars originated in \cite{newton1672letter}, and has steadily evolved since then, with the recognition of absorption lines and spectral features greatly aiding our understanding and classification of stars. As instruments and analysis improved over time, spectroscopy has cemented its place as the method to determine physical properties of stellar atmospheres. Spectroscopic stellar parameters, such as effective temperature $T_\mathrm{eff}$, stellar composition (metallicity, $[M/H]$) and surface gravity ($\log g$), can be determined through an analysis of stellar spectra \citep{gray2005observation}. Spectroscopic parameters have been proven to be useful in deriving reliable masses and radii of stars when combined with evolutionary models \citep{girardi2002theoretical,pietrinferni2004large,dotter2008dartmouth}. The determination of these parameters can therefore provide important data for the understanding and study of stellar evolution \citep[e.g.][]{girardi2000evolutionary}, galactic history \citep[e.g.][]{cunha2016chemical}, star selection for planet detection surveys \citep[e.g.][]{fleming2015apogee},
and characterization of known planet host stars \citep[e.g.]{bean2006metallicities,sweetcat}. Spectroscopy can help to characterize each star's present conditions, as well as letting us infer its past ones and understand their formation environments and galactic populations \citep[e.g.][]{koleva2008spectroscopic,gazzano2010stellar}.

Due to the importance of spectroscopic stellar parameters for the proper characterization of stars, there has been an increase in the number of spectroscopic surveys to characterize FGK stars in the solar neighborhood,  
\citep[e.g.][]{sousa2008spectroscopic,adibekyan2012chemical,tsantaki2013deriving,bensby2014exploring,brewercalifornia}, as well as larger surveys that observe deeper into the Milky Way, such as the Gaia-ESO Survey \citep{gilmore2012gaia}, Galactic Archaeology with HERMES \citep[GALAH,][]{de2015galah}, and the Large sky Area Multi-Object fiber Spectroscopic Telescope \citep[LAMOST,][]{zhao2012lamost}. Most of these studies are focused on spectra observed within visible wavelengths (400-\,nm). However, probing other sections of the electromagnetic spectrum for the derivation of stellar parameters can give us another perspective on stellar characterization through spectroscopy, as well as providing a comparison for stellar parameters derived from optical spectra. The near-infrared wavelengths (NIR), in particular, allows for observations beyond the large molecular clouds existent in the galaxy, as visual extinction is lower in that wavelength range. As FGK stars are still relatively bright in this wavelength range and have features that can be used to characterize them, their NIR spectra can be useful to analyze and characterize these types of stars. Lower temperature stars, like M dwarfs and red giants, have less complex spectra and are relatively brighter in the NIR compared to optical wavelengths, allowing for more precise and accurate spectroscopic parameters \citep[e.g.][]{onehag2012m, rojas2012metallicity}.

We have arrived at the era of instruments that provide high-resolution spectra in the NIR, such as CARMENES \citep[$R=80\,000-100\,000$,][]{quirrenbach2014carmenes}, GIANO \citep[$R\sim50\,000$,][]{origlia2014high}, and SPIROU \citep[$R\sim75\,000$,][]{artigau2014spirou}, as well as large surveys that provide medium-resolution NIR spectra for thousands of stars, like the Apache Point Observatory Galactic Evolution Experiment \citep[APOGEE,][]{prieto2008apogee}. APOGEE is an H-band (1.5-1.7 micron) Sloan Digital Sky Survey program that focuses on obtaining $R \sim 22\,500$ stellar spectra with a 300-fiber spectrograph. It is split between APOGEE-N, using the Sloan 2.5\,m telescope at the Apache Point Observatory in New Mexico \citep{gunn20062}, and APOGEE-S, which uses the 2.5\,m duPont telescope at the Las Campanas Observatory in Chile \citep{bowen1973optical}. It targets mostly red giants and provides public spectra for more than 200\,000 stars in its latest Data Release \citep[DR14,][]{holtzman2018apogee}. In addition to these stars, APOGEE has observed FGK and M dwarfs for calibration purposes or as part of ancillary programs. Parameters for these stars have been derived with APOGEE Stellar Parameter and Chemical Abundances Pipeline \citep[ASPCAP,][]{perez2016aspcap}. ASPCAP works by searching and interpolating a grid of synthetic spectra to find the best match for each observed spectrum, adopting the parameters of the best match as the best parameters for each star. These parameters are then calibrated to best match theoretical models.

In this paper, we provide an alternative methodology to derive spectroscopic parameters using spectral synthesis in the H-band spectra of FGK stars observed by APOGEE and tested its results against similar techniques performed in the optical. Our goal is to have a reliable spectroscopic method for FGK star characterization in the H-band obtained with current and future medium and high resolution H-band spectrographs.

The remainder of this paper is structured as follows: details of our object sample and their literature pertinent to our study are described in Section \ref{data}. The methodology, including the preparation of the APOGEE spectra and all the steps required to obtain stellar parameters from them, are explained in Section \ref{method}.
Section \ref{results} presents both the spectra and parameters derived for solar-type stars, including a visual comparison between the synthetic and the normalized spectra for selected stars. Section \ref{analysis} includes the analysis and comparison of the derived parameters to different literature sources, while possible sources of discrepancies are explored. Finally, in Section \ref{conclusions}, we summarize our conclusions regarding our pipeline and its ability to provide accurate and precise parameters for the Sun and solar-type stars from their APOGEE spectra. 


\section{\label{data} Data}

Our complete sample consists of 3748 stars with H-band spectra from the APOGEE survey Data Release 14 \citep{holtzman2018apogee}. All of these stars have spectroscopic parameters derived with the APOGEE Stellar Parameter and Chemical Abundances Pipeline \citep[ASPCAP][]{perez2016aspcap}. The main goal of ASPCAP is to provide parameters for giant stars, but values for dwarf stars are published as well. The spectroscopic parameters and chemical abundances are determined by ASPCAP in a two step fashion. First, to derive atmospheric parameters such as $T_\mathrm{eff}$, $[M/H]$, and $\log g$, APOGEE observations are compared to a large library of synthetic spectra to find the spectra that best matched the observed one by interpolating different synthetic models. This library is separated into five smaller sections, for GK dwarfs, GK giants, M dwarfs, M giants, and F-type stars. Secondly, to derive the abundance of individual elements, the atmospheric parameters obtained from the best fit synthetic spectra are used to fit limited regions of the spectra dominated by spectral features associated with each given element. Afterwards, $T_\mathrm{eff}$ and $[M/H]$ values for dwarf stars are both calibrated using independent methods to ensure their accuracy. $T_\mathrm{eff}$ values were calibrated by minimization of the differences between ASPCAP and photometric observations by \cite{hernandez2009new}. $[M/H]$ values were internally calibrated as a function of $T_\mathrm{eff}$. In addition, a zero-point shift was adopted to force the mean abundance ratios of all observed stars with solar metallicity to be zero. Despite finding their $\log g$ values for M dwarfs to be too low based on expectations from stellar isochrones, due to a lack of a significant number of asteroseismic calibrators available for dwarf stars, ASPCAP does not calibrate them in any way.\footnote{Exact values and calibrations are available in \citep{holtzman2018apogee}, and more information can also be consulted at the APOGEE website https://www.sdss.org/dr14/irspec/aspcap/.}

\begin{figure}
\resizebox{\hsize}{!}{\includegraphics{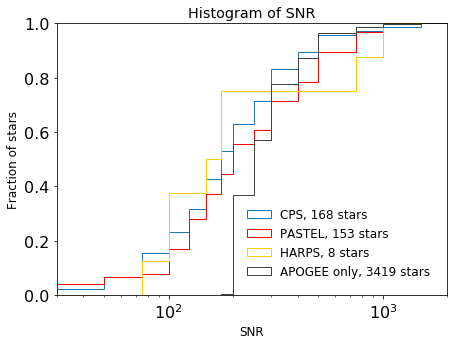}}
\caption{Cumulative histogram of the APOGEE-published $S/N$ of our stellar sample.}
\label{snr}
\end{figure}

Our selection was done by selecting objects with ASPCAP $T_\mathrm{eff}$ values between $5500-6200$\,K, $S/N\geq200$ and $-0.5\,$dex$<[M/H]<0.5\,$dex. This was done to exclude stars with low-quality spectra and to ensure a sample both homogeneous and with parameters close to solar. A total of 3419 main sequence stars matching these criteria are in the sample. Additional stars, not meeting these criteria, were included as well as part of subsamples previously observed in the optical. Most of these additional stars are also main-sequence FGK stars, but some of them are turn-off stars, and have therefore lower $\log g$ values than the ones still in the main-sequence. These subgiant stars were included for statistical reasons, as the number of FGK main-sequence stars in common between APOGEE and other optical surveys is not large enough for our purposes, and to test the limits of  the parameter space. Our primary focus remains on the main-sequence stars. The total of 3748 stars includes those additional stars observed in optical wavelengths. Those stars are described in section \ref{subsample}. The APOGEE $S/N$ distribution of our sample of spectra is shown in Fig. \ref{snr}.

\subsection{Sample characterization}

\begin{figure}
\resizebox{\hsize}{!}{\includegraphics{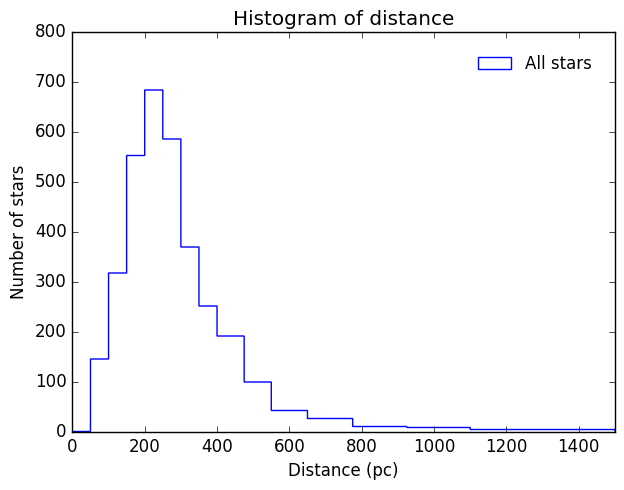}}
\caption{Histogram of the calculated distances to the sample stars, using Gaia Data Release 2 values for the parallax \citep{brown2018gaia}.}
\label{distance}
\end{figure}

\begin{figure}
\resizebox{\hsize}{!}{\includegraphics{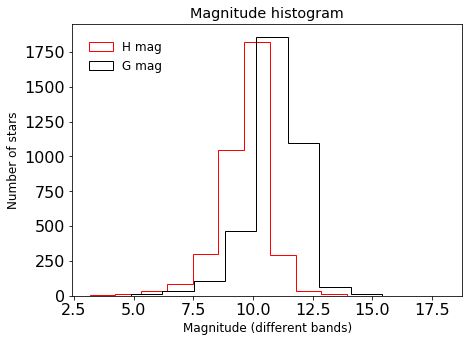}}
\caption{Histogram of the magnitudes for stars in the sample. \textit{G} band data comes from \cite{brown2018gaia}; \textit{H} band from APOGEE.}
\label{magnitude}
\end{figure}

\begin{figure}
\resizebox{\hsize}{!}{\includegraphics{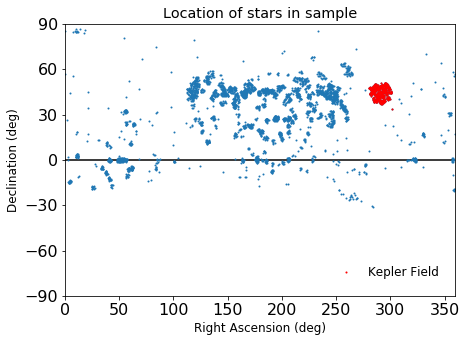}}
\caption{Map showing the location in the sky (Right Ascension and Declination) of our sample stars.}
\label{declination}
\end{figure}

Parallaxes, positions, and optical photometry for 3645 stars in our sample were found in the Gaia Data Release 2 \citep{brown2018gaia}. Due to either inaccurate proper motion measurements or lack of observations from Gaia, no matches were discovered for the remaining 103 stars, but their overall distribution should match that of the other stars. The large majority of the stars are located between 100\,pc and 500\,pc from us, with only a few reaching distances above 1\,Kpc (see Fig. \ref{distance}). Therefore, our sample, while not entirely composed of solar neighborhood stars ($d<50\,pc$), consists of local stars. Their magnitudes in the Gaia and H-band filters are shown in Fig. \ref{magnitude}. Most of these objects exhibit magnitudes in the ranges of 9 < \textit{G} < 13 and 8 < \textit{H} < 11. Their distances and magnitudes confirm to us that the sample is composed of FGK dwarf stars, as any giant stars in the sample would either be brighter or further away from us. Their location in the sky is shown in Fig. \ref{declination}. Given that the majority of the observations belong to APOGEE-N, it is no surprise that most of the stars are located in the northern celestial hemisphere. The concentric circles represent the fields chosen by the APOGEE team for follow-up. In red, 449 stars in our sample that are a part of the Kepler Field \citep{latham2005kepler} are shown.

\subsection{\label{subsample} Comparison sub-samples}

We selected three works in the literature that have analyzed APOGEE stars with medium or high resolution optical spectra. These stars were added to our full sample, even if they did not meet our selection criteria ($5500\,K < T_\mathrm{eff} < 6200\,$K, $S/N\geq200$ and $-0.5$\,dex$<[M/H]<0.5\,$dex). This was done so more stars with accurate stellar parameters were available for the comparison of our results.

The California Planet Survey \citep[][hereby referred to as CPS]{brewercalifornia} used data from the HIRES spectrograph ($R \sim 70000$)at the Keck Observatory. The stellar parameters were obtained with the semi-automated procedure SME \citep{valenti1996spectroscopy}, where the observed spectra were fitted iteratively with synthetic spectra from 1D local thermodynamic equilibrium (LTE) plane-parallel MARCS atmosphere models. CPS provides rotational velocities and abundances for 15 elements as well. The reported precisions are of 25\,K for $T_\mathrm{eff}$, 0.01\,dex for $[M/H]$ and 0.028\,dex for $\log g$. One-hundred and sixty eight stars in our sample are found in the CPS catalog.

The PASTEL catalog \citep{soubiran2016pastel} is a compilation of spectroscopic parameters from high-resolution ($R \geq 25 000$) spectra with $S/N\geq 50$. It contains results from different sources in the literature that derived stellar parameters with model atmospheres. The reported uncertainties vary, with median errors of $\sim 1.1\%$ for the $T_\mathrm{eff}$ ($\sim65\,$K for a sun-like star), $\sim0.06\,$dex for $[Fe/H]$ and $0.10\,$dex for $\log g$. A total of 157 stars in our sample are found in the PASTEL catalog. However, as 4 of these stars are also found in CPS, only 153 stars were considered as the PASTEL comparison sample.

\cite{harps2017mena} revised the spectroscopic parameters and abundances of 1111 FGK dwarf stars of the High Accuracy Radial velocity Planet Searcher Guaranteed Time Observations planet search program \citep[HARPS-GTO, ][]{pepe2000harps,pepe2011harps}. The EW method was used on HARPS-GTO $R\sim115 000$ spectra for parameter and abundance determinations \citep{sousa2008spectroscopic, tsantaki2013deriving}. Their average cited internal errors are 24\,K for $T_\mathrm{eff}$, 0.02\,dex for $[M/H]$ and 0.03\,dex for $\log g$. Given that the HARPS-GTO planet survey concentrates on bright inactive stars (mostly $V_{mag}<11$), only 8 stars of our sample are part of the 1111-star sample characterized in \cite{harps2017mena}.

\section{\label{method} Methodology}

This section describes our method and the steps required to derive parameters from APOGEE's H-band spectra.

\subsection{\label{ispec}\textit{iSpec} and \textit{Turbospectrum}}

\textit{iSpec} is a multi-purpose python-based tool designed to derive atmospheric parameters from stellar spectra through different methods \citep{blanco2014determining,ispec2019sbc}. Among the radiative transfer codes available to use within \textit{iSpec}, we chose for spectral synthesis \textit{Turbospectrum}, developed by \citet{plez1998} and \citet{plez2012turbospectrum}. \textit{Turbospectrum} is the fastest code available within \textit{iSpec} and has shown to be compatible with our line list.

The code uses a least-squares algorithm to match the two spectra, using the synthetic spectra generator chosen by the user. The code will then run its minimization routine and provides an output that corresponds to the spectra that best matches the input. It provides an estimate of the errors, based on the spectrum errors, as well as the $\chi^2$ value for the best match. These values are included along with our derived parameters for the stars in Table \ref{table:CPSresults}.

\subsection{Stellar models}

\textit{iSpec} is compatible with two categories of models available: \textit{MARCS} \citep{MARCS} and \textit{ATLAS9} \citep{atlas9}. There are some subcategories of models within each of these. We chose the \textit{MARCS.GES} models for our syntheses, since it offers better coverage in the parameter space we are interested in. The \textit{MARCS.GES} models cover $T_\mathrm{eff}$ from 2500\,K to 8000\,K and $\log g$ from 0.0 dex to 5.5\,dex, allowing us to extend our method to the M dwarf regime using the same models (Sarmento et al. in prep). The \textit{MARCS.GES} models assume plane-parallel 1D stratification, hydrostatic equilibrium, mixing-length convection, and local thermodynamic equilibrium. \textit{MARCS.GES} also take into account $[\alpha/Fe]$ and $[O/Fe]$ parameters when generating the models. Standard models for $[Fe/H]<-0.25\,$dex have $[\alpha/Fe]$ and $[O/Fe]$ values tailored to match abundances measured in the solar neighborhood stars.

For the solar abundances, the values used were the ones available in \cite{grevesse2007solar}, as they were the ones used to create the MARCS models used in the synthesis.

\subsection{\label{Norm} Normalization}

APOGEE DR14 spectra includes both normalized spectra and multiple stages of spectrum processing for its stars \footnote{https://www.sdss.org/dr14/data\_access/}. The normalized spectra are processed by ASPCAP to derive parameters and could, in principle, be used to fit individual lines and determine abundances and parameters using the EW method, for example. However, the normalization done for APOGEE DR14 spectra is not precise enough to compare the normalized spectrum to our full synthetic spectrum. Therefore, we had to perform our own normalization to the combined spectra \footnote{https://data.sdss.org/sas/dr14/apogee/spectro/redux/r8/stars/} of each star in our sample.

Our normalization method relies on what we call a reference spectrum. We created a grid of normalized synthetic spectra covering the expected parameter space of our sample of stars with \textit{iSpec}, \textit{Turbospectrum} and \textit{MARCS.GES} models. The $T_\mathrm{eff}$ values of the grid are 4200, 4600, 5000, 5400, 5700, 6000, 6400\,K, so the grid has a slightly smaller step near the solar $T_\mathrm{eff}$. It covers $[M/H]$ from -1.6\,dex to 0.4\,dex and $\log g$ from 4.2\,dex to 5.4\,dex (both in steps of 0.4\,dex). We know that some of our stars are in the subgiant branch and the $\log g$ values could be lower in those cases in order to more accurately match their spectra. The normalization grid properly normalizes areas of the spectra with large absorption lines. The smoothing after the normalization erases information from the weaker features, making it possible to use the grid to normalize spectra that does not match exactly the same parameters but that falls relatively close. Since the characterization of these subgiants is not the main focus of the work, we limited the parameter space of our normalizations. The most similar synthetic spectrum to each observed star, taking all three parameters into account, is selected as its reference spectrum. The parameters used for the selection are from ASPCAP or from literature sources when available (PASTEL, CPS, or HARPS-GTO).

The normalization is done by dividing the observed spectrum by its reference spectrum. The divided spectrum is smoothed afterward, taking the median value within 5.0\,nm for each point in the spectrum. Dividing the observed spectra by this shape-fit results in a normalized spectrum. Our normalization method preserves the wide absorption features in the normalized spectrum, such as the hydrogen Brackett series lines ($1736.69$\,nm, $1681.11$\,nm, $1641.17$\,nm, $1611.37$\,nm), that the default APOGEE normalization misses. The lines around $1681.11$\,nm, $1641.17$\,nm, however, were removed from the line mask as synthesizing them with the necessary accuracy for parameter determination proved to be impossible.

\subsection{\label{linelist} Line-list}

The line-list used is a compilation of lines from two different sources: 
The Vienna Atomic Line Database \citep[VALD,][]{piskunov1995vald} and the APOGEE line-list \citep{shetrone2015sdss}\footnote{The strength ($\log(gf)$) of the lines in this list was adjusted to better match both a Solar and an Arcturus synthetic spectra. We did not change their values, which might have slightly affected our results.}. They contain all of the relevant elemental lines as well as the molecular lines for CO, OH, $\textrm{C}_{2}$, CN, CH, and FeH. The solar spectrum presented in \cite{wallace1996infrared} was used to validate the final line-list. Two synthetic spectra were created for the Sun, one with the VALD line-list and another with the APOGEE line-list. All the spectra, synthetic and solar, were convolved to the APOGEE-N resolution. Then, individual 0.2\,nm regions in each synthetic spectrum were compared to the spectrum from the Sun, to inspect which line-list provided the best fit for each region. Our line-list is then composed by the best-fitting lines to the solar spectra from the VALD and APOGEE line-lists and contains 85334 lines.

\subsection{\label{linemask} Line masks}

\textit{iSpec} can use line masks to select the spectral features that are going to be considered for the determination of the atmospheric parameters. To create a list of line masks covering the best synthetic absorption lines, we compared a synthetic spectrum ($T_\mathrm{eff}$ 5777\,K, $[M/H]$ 0.0 and $\log g=4.44$\,dex) to the solar spectrum by \cite{wallace1996infrared}, both at APOGEE’s resolution. The lines were selected by excluding 1) regions with lines that did not appear in the synthetic spectrum and 2) regions in the synthetic spectrum that did not match the solar spectrum as expected.

The inspection of each region in the synthesized and observed spectrum was made visually (see Fig. \ref{solarspectra}). The same line mask was used in all stars in our sample.

\subsection{\label{error_est} $\chi^2$ fit and error estimation}

The $\textit{mpfit.py}$ code, based on MPFIT \citep{markwardt2009non}, is used by \textit{iSpec} to minimize the $\chi^2$ difference between the synthetic and the observed spectra. It uses a Levenberg-Marquardt least-squares minimization to obtain the best fit to a given spectrum considering only the regions
that are included in the line mask described above. All the errors cited for the stellar parameters in this work were calculated by \textit{iSpec} from this code, and correspond to the formal $1-\sigma$ errors of each parameter, computed from the covariance matrix. The error calculation assumes uncertainties in the observed
spectrum equal to the flux divided by the $S/N$ values published by ASPCAP. These are considered as internal errors and are reported in each parameter for all stars in Table \ref{table:CPSresults}.

To address how well our method is able to recover the parameters of each star under slightly different conditions, we added random Gaussian noise based on the flux errors to the normalized spectrum for each star. The Gaussian noise added had a zero mean and a standard deviation equal to the estimated flux error for each individual pixel. To avoid individual points or areas with large reported flux errors (>1/10th of the pixel flux value) that could skew the analysis, we restricted the value of Gaussian noise standard deviation to 1/10th of the flux for those pixels.

\begin{figure}
\resizebox{\hsize}{!}{\includegraphics{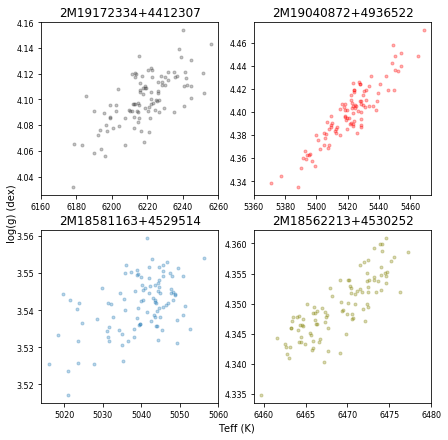}}
\caption{Comparison between output parameters ($\log g$ and $T_\mathrm{eff}$) of 4 selected stars in 100 different iterations. Each point in the spectra had random Gaussian noise added to it.}
\label{scrambledlogg}
\end{figure}

\begin{figure}
\resizebox{\hsize}{!}{\includegraphics{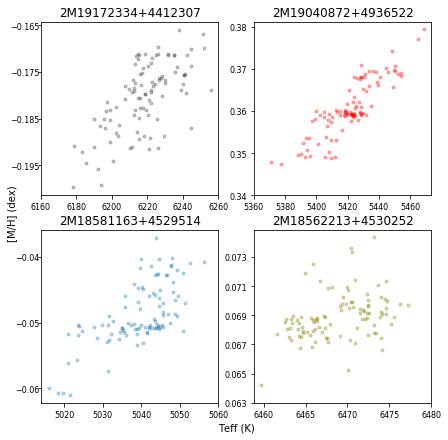}}
\caption{Comparison between output parameters ($[M/H]$ and $T_\mathrm{eff}$) of 4 selected stars in 100 different iterations. Each point in the spectra had random Gaussian noise added to it.}
\label{scrambledmh}
\end{figure}

We chose 4 stars with different spectral derived parameters from the CPS catalog and different $S/N$ on their APOGEE spectra and added to each spectrum the random Gaussian noise described above 100 times for each star. The derived parameters by \textit{iSpec} from the 100 random Gaussian noise-added spectra are shown in Figs.\ref{scrambledlogg} and \ref{scrambledmh}. These figures also show the strong degeneracies between parameters, as increases in derived $T_\mathrm{eff}$ correspond to increases in $\log g$ and $[M/H]$ as well. Table \ref{table:scrambled_stars} shows the mean values and standard deviation derived from the 100 iterations for each star. The consistency of our method across multiple syntheses can be confirmed, being able to recover consistent values for all the stars. As expected, the stars with the lowest $S/N$, \object{2M19172334+4412307} and \object{2M19040872+4936522}, exhibit the largest differences between syntheses, reaching up to $\Delta T_\mathrm{eff} \sim 100\,$K and $\Delta \log g = 0.11$\,dex.

\begin{table*}
\caption{Parameters derived in the 4 stars selected. The C indicates the literature (CPS) parameter for that star, and $\overline{parameter}$ and $\sigma$ indicate the average value and the standard deviation measured across all 100 iterations, respectively.}
\label{table:scrambled_stars}
\centering
\begin{tabular}{c c c c c c c c}
\hline \hline
APOGEE\_ID & $S/N$ & $\overline{T_\mathrm{eff}}\pm\sigma _{T_\mathrm{eff}}$  & $T_\mathrm{eff}^{C}$ & $\overline{\log g}\pm\sigma _{\log g}$ & $\log g^{C}$ & $\overline{[M/H]}\pm\sigma _{[M/H]}$ & $[M/H]^{C}$\\ 
\hline
\object{2M19172334+4412307 }   &    121    & $    6218    \pm    16    $\,K & 6221\,K & $    4.10    \pm    0.02    $\,dex & 4.13\,dex & $    -0.18    \pm    0.01    $\,dex &  -0.14\,dex  \\
\object{2M19040872+4936522 }  &    132    & $    5421   \pm    18    $\,K & 5487\,K & $    4.40   \pm    0.03    $\,dex & 4.31\,dex & $    0.36    \pm    0.01    $\,dex & 0.34\,dex \\
\object{2M18581163+4529514 }   &    216    & $    5039   \pm    9    $\,K & 4989\,K & $    3.54    \pm    0.01   $\,dex & 3.37\,dex & $    -0.05    \pm    0.01    $\,dex  & -0.04\,dex  \\
\object{2M18562213+4530252 }   &    1095    & $    6468    \pm    6   $\,K & 6378\,K & $    4.35   \pm    0.01    $\,dex & 4.2\,dex & $    0.07    \pm    0.01    $\,dex & 0.18\,dex \\
\hline
\end{tabular}
\end{table*}

\subsection{\label{free} Free and fixed parameters within iSpec}

\textit{iSpec} allows for the customization of the free parameters in each analysis during the $\chi^2$ fit and error minimization processes. All syntheses done for this work had the same group of free parameters: effective temperature ($T_\mathrm{eff}$), surface gravity ($\log g$), metallicity ($[M/H]$), micro-turbulent velocity ($v_{mic}$), projected rotational velocity ($v \sin i$) and spectral resolution. The input values were kept at 5800\,K for the $T_\mathrm{eff}$, 0.0\,dex for the $[M/H]$, 4.5\,dex for the $\log g$, 1.06\,km/s for the $v_{mic}$, 1.6\,km/s for $v \sin i$, and  22000 for the resolution. The macro-turbulent velocity ($v_{mac}$) follows an empirical relation provided by \textit{iSpec} and built by the Gaia-ESO Survey based on their dataset and the $T_\mathrm{eff}$, $\log g$ and $[M/H]$ of the stars. The limb darkening is a required parameter for syntheses using iSpec. We tested different input values compatible with solar-type stars and since the output results have a negligible variation we decided to fix it at 0.6.

\section{\label{results} Results}

\subsection{\label{solar} Solar spectra}

Despite it previously being used to select lines for our line-list, we retrieved the solar spectra using our method in order to confirm its reliability. Since the Sun is the closest and best-characterized star available, any reliable synthesis-based method should recover its spectral characteristics. Therefore, to calibrate and test \textit{iSpec} with \textit{Turbospectrum} in the H-band, a solar spectrum was synthesized using the methodology described in section \ref{method}. We used the solar spectrum by \cite{wallace1996infrared} observed with the Fourier transform spectrometer at the Math-Pierce solar telescope on Kitt Peak, with a resolution of $300\,000$ and high $S/N$. The spectrum was degraded down to $R = 22\,500$ to match APOGEE's resolution. It was normalized using the template method described in section \ref{Norm} with a reference spectrum of $T_\mathrm{eff} = 5700$\,K, $[M/H] = 0.0$\,dex and $\log g = 4.5$\,dex.

\begin{landscape}

\begin{figure}
\resizebox{\hsize}{!}{\includegraphics{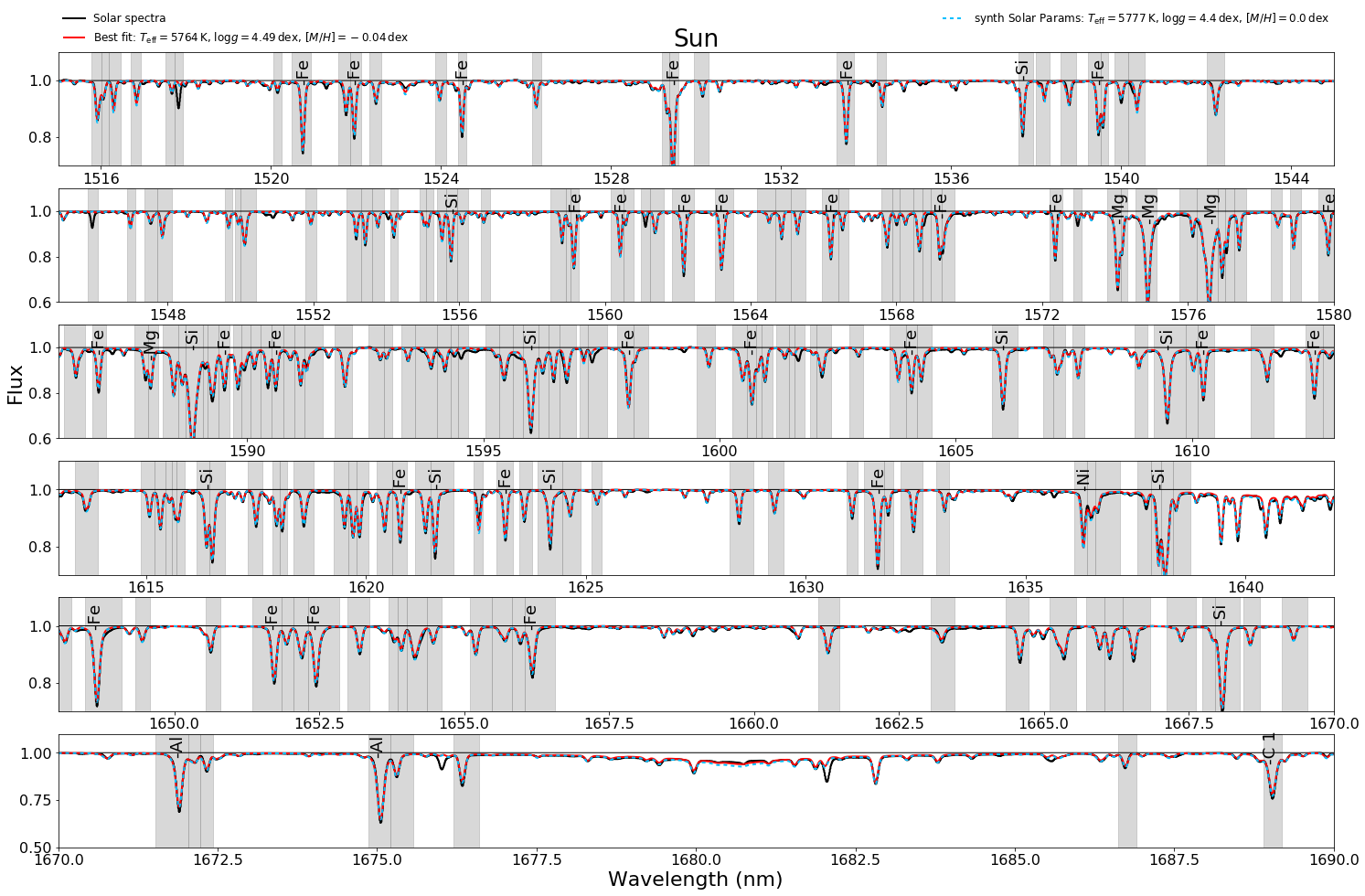}}
\caption{Comparison between Solar spectra (black), the best fit from our pipeline synthetic spectra (red), and a second synthetic spectrum made with the standard solar parameters (blue) for APOGEE wavelength range. 
The parameters derived were $T_\mathrm{eff}=5764\pm45$\,K,$\log g=4.49\pm0.09$\,dex, $[M/H] = -0.04\pm0.03$\,dex, compared to the standard ones of $T_\mathrm{eff}=5777$\,K,$\log g=4.4$\,dex, $[M/H] = 0.0$\,dex.
The gray areas indicate the regions of the spectra included in the line masks (see section \ref{linemask}).}
\label{solarspectra}
\end{figure}

\end{landscape}

\begin{figure}
\resizebox{\hsize}{!}{\includegraphics{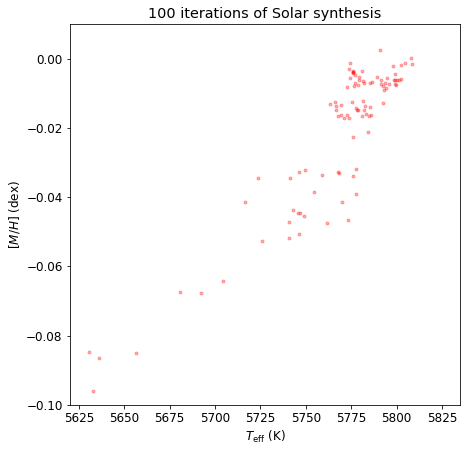}}
\caption{Output $T_\mathrm{eff}$ and $[M/H]$ across 100 iterations of our pipeline with the solar spectra.}
\label{solar100}
\end{figure}

The observed and synthetic solar spectra are shown in Fig. \ref{solarspectra}. We took as the accepted solar parameters $T_\mathrm{eff} = 5777$\,K, $\log g = 4.44$\,dex and $[M/H] = 0.0$\,dex \footnote{https://nssdc.gsfc.nasa.gov/planetary/factsheet/sunfact.html}. The synthesis done with our methodology provides a synthetic spectrum that matches the solar spectrum and derives just slightly lower values for the solar spectra at APOGEE's resolution: $T_\mathrm{eff}=5764\pm35$\,K,$\log g=4.49\pm0.07$\,dex, $[M/H] = -0.04\pm0.02$\,dex. Given the \cite{wallace1996infrared} solar spectrum did not provide flux errors, we estimated errors for the parameters from 100 syntheses of the solar spectrum with injected Gaussian noise of 1/1000th of the flux at each pixel of the spectra (similar to the process described in section \ref{error_est}). The average values for $T_\mathrm{eff}$, $\log g$, and $[M/H]$ from the 100 syntheses are shown in Table \ref{table:solar_100}. As expected, the parameters obtained in each synthesis are not independent of each other, and there are degeneracies between them, as shown by the correlation between $T_\mathrm{eff}$ and $[M/H]$ in Fig. \ref{solar100}. Within the 100 syntheses, 87 have estimated $T_\mathrm{eff}$ values within $5777\pm50\,$K, and 65 of them have $[M/H]$ within $0.0\pm0.02$\,dex. These results also show standard deviations of 45\,K for $T_\mathrm{eff}$, 0.09\,dex for $\log g$, and 0.03\,dex for $[M/H]$. The parameter analysis indicates a high degree of consistency between the iterations of our pipeline. However, in some cases, the retrieved parameters diverge significantly from the expected. In particular, we registered unusual differences in $T_\mathrm{eff}$ of $-150\,$K and -0.10\,dex for $[M/H]$.

\begin{table}
\caption{Solar parameters derived from \cite{wallace1996infrared} solar spectrum, and average results across 100 iterations with injected errors.}
\label{table:solar_100}
\centering
\begin{tabular}{c c c c}
\hline
Parameter & Output & Average 100 iter. & $\sigma$ \\ 
\hline
$T_\mathrm{eff}$(K) & 5764 & 5763 & 45 \\
$\log g$(dex)  & 4.49 & 4.51 & 0.09 \\
$[M/H]$(dex) & --0.04 & --0.02 & 0.03 \\
\hline
\end{tabular}
\end{table}

\subsection{\label{FGK_results} APOGEE FGK sample}

\begin{figure}
\resizebox{\hsize}{!}{\includegraphics{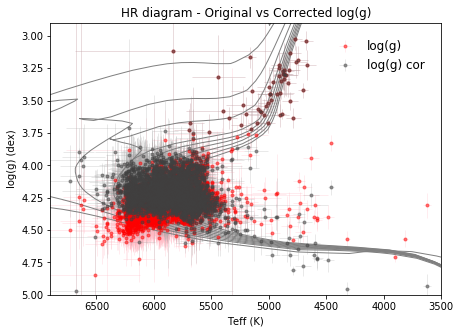}}
\caption{Pseudo-HR diagram of our results with the full sample, plotting both the derived $\log g$ and $\log g$ after corrections based on trigonometric $\log g$ from \cite{harps2017mena} against $T_\mathrm{eff}$
for all the stars in the sample. Worst outlier stars are excluded from this plot. PARSEC isochrones with $[M/H]=0.0\,$dex from \citet{bressan2012parsec} are overplotted.}
\label{HR_diagram}
\end{figure}

The 3748 stars in our sample were synthesized using the methodology described in Section \ref{method}. Fig. \ref{HR_diagram} shows in red the distribution of $\log g$ vs $T_\mathrm{eff}$ values for our sample stars, with overplotted isochrones from PARSEC \citep{bressan2012parsec} made with $[M/H]=0.0\,$dex. As expected from our selection of objects described in section \ref{data}, the large majority of the stars synthesized fall within a small range of parameters, corresponding to FGK main-sequence stars. They also mostly fall within the range of the isochrones, giving us more confidence on the values. A small number of stars exhibit lower temperatures than 5\,000\,K. These stars also appear as subgiants in the CPS sample ($\log g < 3.9$\,dex) and as cooler dwarfs in the PASTEL catalog. The $\log g$ for these colder stars seems to be slightly underestimated, as they are outside the range of the isochrones. Moreover, the $\log g$ for some of the hottest stars in the sample seems to be overestimated. A similar behavior for $\log g$ values of FGK dwarfs in the HR diagram has been observed after using the ionization balance method to derive $\log g$. Therefore, we decided to apply the corrections based on trigonometric $\log g$ derived in \cite{harps2017mena} (see Equations 1,2 and 3 in that paper) to $\log g$ values above 3.75\,dex. Our corrected $\log g$ values are also plotted in gray in Fig.\ref{HR_diagram} and are included as $\log g_{cor}$ in Table \ref{table:CPSresults}.

\begin{figure}
\resizebox{\hsize}{!}{\includegraphics{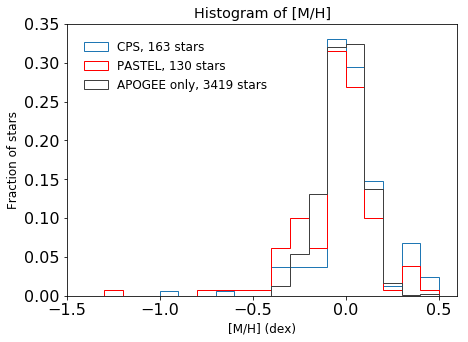}}
\caption{Histogram with the metallicity distribution of the final sample, divided by subsamples of stars in common with other catalogs. Values plotted are the output of our pipeline. Worst outlier stars are excluded from this histogram.}
\label{MH_plot}
\end{figure}

The $[M/H]$ histograms for each significant subsample are shown in Fig. \ref{MH_plot}. The $[M/H]$ values by \textit{iSpec} follow a Gaussian distribution that peaks around 0.0\,dex, which is the Sun's metallicity. This was expected for our APOGEE-only subsample, since we restricted the sample by metallicity ($-0.5<[M/H]<+0.5\,$dex only). Analysis of metallicities in the solar neighborhood sample \citep[as seen, for example, in][]{sousa2008spectroscopic} show a slightly lower average value of $[Fe/H]=-0.09$.

To reveal the challenges of synthesizing APOGEE spectra, both the observed APOGEE and synthesized spectra for an example star is shown in Fig.\ref{PASTEL_Gstar}\footnote{Comparison spectra for 3 additional stars is available in the appendix.} This star was chosen as a representatives of the sample and of the output produced by the method, both the synthetic spectrum generated and the parameters derived by the pipeline. For the example star, the APOGEE spectrum is shown in black, the 'Best fit' spectrum (synthetic spectrum created with the parameters found by our methodology) is shown in red, and the synthetic spectrum created with the full line-list and using the literature parameters is shown in blue. Therefore, the blue spectrum represents the expected spectrum for the example star considering its derived parameters from high-resolution optical spectra.

The star \object{2M19144528+4109042} (\object{KOI-85}, Fig. \ref{PASTEL_Gstar}) has 3 sub-stellar companions detected \citep{borucki2011characteristics,rowe2014validation}. Our spectroscopic values are higher than the optical ones by the CPS catalog, except for the metallicity, which is slighter lower. However, considering errors, the values derived by us and from the CPS catalogs are consistent. Both synthetic spectra are similar between themselves and are a match to the observed spectrum; the most significant differences appear around the strong hydrogen line in the 1681.1 nm region and continuum around 1640\,nm. There are also differences in line depths across the spectrum, most noticeably the Si line at wavelength 1596\,nm. This can be explained by the differences in metallicity between the two synthetic spectra.

We realize that comparing the best-fit synthetic spectra with ones created using literature parameters is not a perfect comparison due to inhomogeneities in the literature parameters, as they were derived with different spectra, methods, models, and normalizations than the ones we use. Using optical spectra to derive parameters for our sample stars and using them to generate NIR spectra could result in a better comparison for our method. However, different line-lists and masks would still be required for the derivation of parameters with optical wavelength spectra, which can result in differences when comparing them with H-band results. The comparisons present in this paper are therefore a compromise between these issues.

We are also aware that the visual comparisons between the spectrum presented in this paper are difficult to judge accurately, as the minimization process takes into account flux errors that are not shown here. The spectrum shown in Fig.\ref{PASTEL_Gstar} is intended more as a visual display of the method's results than its scientific precision.

\begin{landscape}

\begin{figure}
\resizebox{\hsize}{!}{\includegraphics{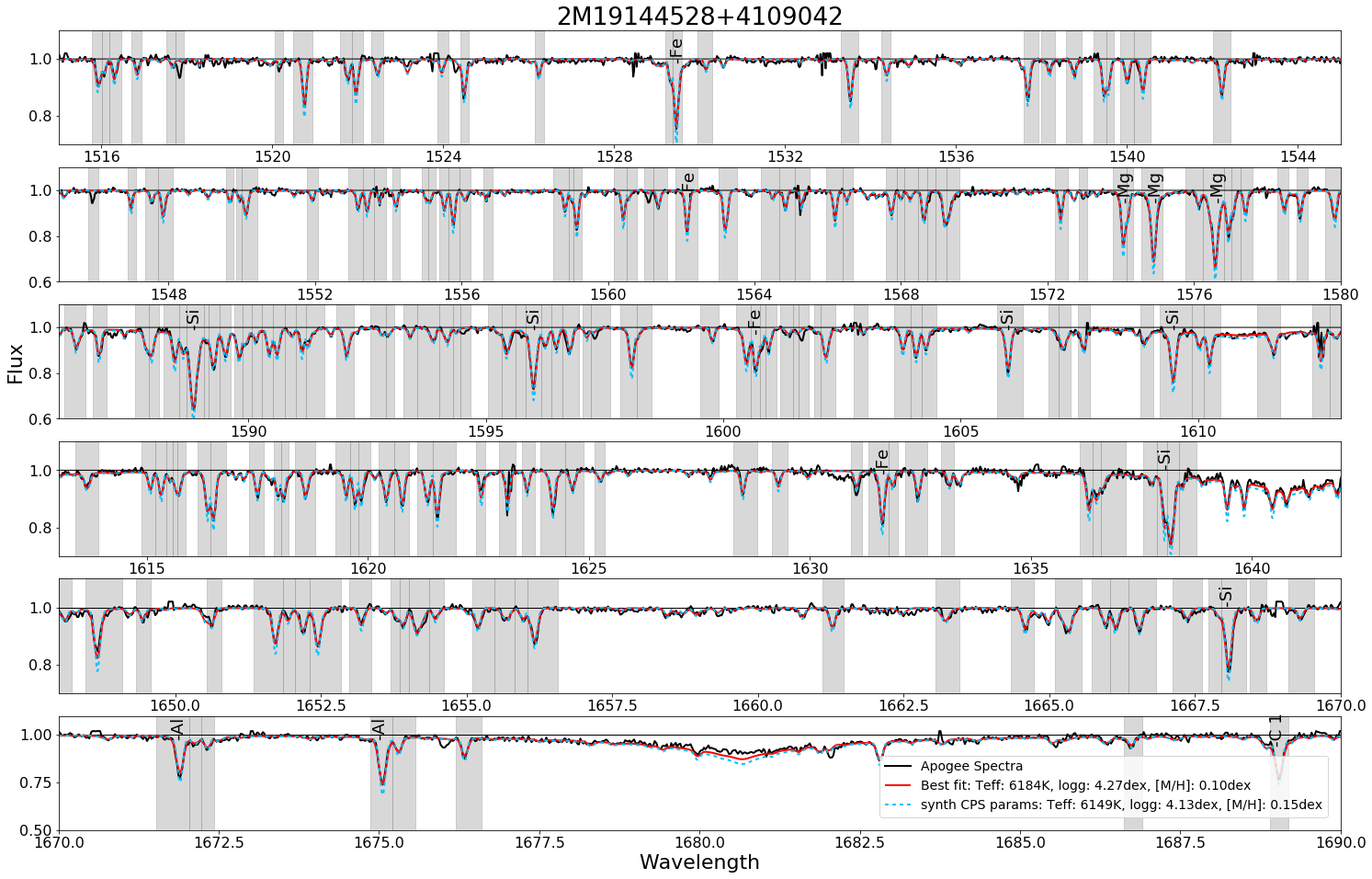}}
\caption{Comparison between APOGEE spectra of the star \object{2M19144528+4109042} (\object{KOI-85}, black) and two synthetic spectra (red, straight line, 
with the our pipeline's best match parameters and blue, dashed, with the CPS parameters for this star) for APOGEE wavelength range. 
In gray highlight are the areas used for $\chi ^2 $ minimization by our pipeline's algorithm.
The best match parameters derived were $T_\mathrm{eff}=6195\pm164$\,K,$\log g=4.28\pm0.21$\,dex, $[M/H] = +0.11\pm0.06$\,dex, 
and the ones published in CPS were $T_\mathrm{eff}=6149\pm25$\,K,$\log g=4.13\pm0.028$\,dex, $[M/H] = +0.15\pm0.01$\,dex.}
\label{PASTEL_Gstar}
\end{figure}

\end{landscape}

\section{ \label{analysis}Discussion}

\subsection{\label{ASPCAP} ASPCAP comparison}

\begin{figure}
\resizebox{\hsize}{!}{\includegraphics{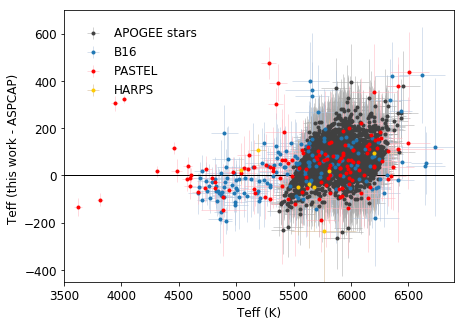}}
\caption{Diagram representing the output $T_\mathrm{eff}$ from our pipeline and the difference between it and the values published by ASPCAP for each star in our sample. Only ASPCAP values are being compared and the different colors are used to distinguish between stellar subsamples. Errors displayed in both axis are our pipeline's estimates using the covariance matrix.}
\label{aspcap_teff}
\end{figure}

\begin{figure}
\resizebox{\hsize}{!}{\includegraphics{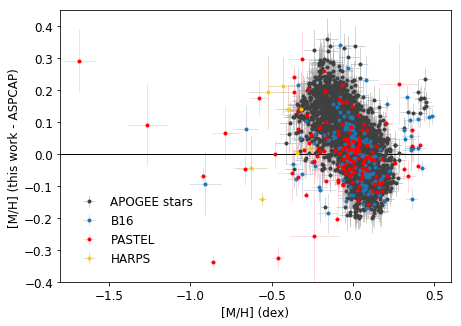}}
\caption{Diagram representing the output $[M/H]$ from our pipeline and the difference between it and the values published by ASPCAP for each star in our sample. Only ASPCAP values are being compared and the different colors are used to distinguish between stellar subsamples. Errors displayed in both axis are our pipeline's estimates using the covariance matrix.}
\label{aspcap_mh}
\end{figure}

\begin{table}
\caption{Median differences for ASPCAP parameters. The 'Stars' column represents the number of stars with this ASPCAP parameter between the two original samples}
\label{table:aspcap}
\centering
\begin{tabular}{c c c}
\hline
Parameter & Stars & Median (This Work - ASPCAP)  \\ 
\hline
$T_\mathrm{eff}$ & 3670 & +62\,K \\
$[M/H]$ & 3668 & +0.04\,dex \\
$\log g$ & 3622 & -0.01\,dex\\
\hline
\end{tabular}
\end{table}

In this subsection, we compare our results to the derived ASPCAP parameters for our sample of stars. ASPCAP was developed to derive spectroscopic parameters for the APOGEE survey, and its parameter determination pipeline is optimized for giant stars. As mentioned in Section \ref{data}, $[M/H]$, $T_\mathrm{eff}$, and $\log g$ are provided for all stars in the sample. ASPCAP values for $[M/H]$ and $T_\mathrm{eff}$ for all stars are calibrated using independent methods, but the $\log g$ values for dwarf stars are not independently calibrated and differ from isochrone values. We decided to compare our values for $T_\mathrm{eff}$ and $[M/H]$  with the calibrated ones provided by ASPCAP for our sample of stars. We also compare our raw values for $\log g$ with the available ASPCAP $\log g$ values.

Fig. \ref{aspcap_teff} presents the $T_\mathrm{eff}$ comparison, while the $[M/H]$ comparison is presented in Fig. \ref{aspcap_mh}. These comparisons include both the subsamples characterized in optical wavelengths and the 3419 stars from which only ASPCAP parameters are available. Despite only ASPCAP values being compared, we distinguish the subsamples characterized in the optical by color (CPS - blue; PASTEL - red; HARPS-GTO - yellow), while the stars in black have only been characterized by ASPCAP and our pipeline. This is done for visual clarity and consistency reasons.

In Table \ref{table:aspcap}, the average differences between our output and the ASPCAP values are presented. With a sample of 3419 stars being so large when compared to our subsamples of stars observed in the optical, the fact that the measured differences between our pipeline's values and ASPCAP's parameters are within one standard deviation indicates that our method can provide consistent parameters for solar-type stars in the NIR.

The median estimated $T_\mathrm{eff}$ is 62\,K above ASPCAP values, although the difference varies between --150\,K and +300\,K. The fact that the median difference is still within the margin of error for both pipelines is an encouragement for the accuracy of our method.

A small difference is present in metallicity as well, which can clearly be seen in Fig. \ref{aspcap_mh} and in Table \ref{table:aspcap}. We find an average difference between our values for $[M/H]$ and the ones in ASPCAP of +0.04\,dex. Comparing with the differences registered against parameters measured in the optical, we find this value to be above the median differences for the CPS values (-0.02\,dex) and at the same level as the PASTEL values (+0.04\,dex, see Section \ref{litcomp}). It is still below the HARPS-GTO difference of +0.16\,dex. Additionally, there seems to be a trend, observable in Fig. \ref{aspcap_mh}, that results in lower $\Delta [M/H]$ values for stars with higher $[M/H]$, with a small number of high $[M/H]$ outliers. A larger sample size of both low and high-$[M/H]$ stars seems necessary to better characterize the behavior of the pipeline for that extended parameter space.

\begin{figure}
\resizebox{\hsize}{!}{\includegraphics{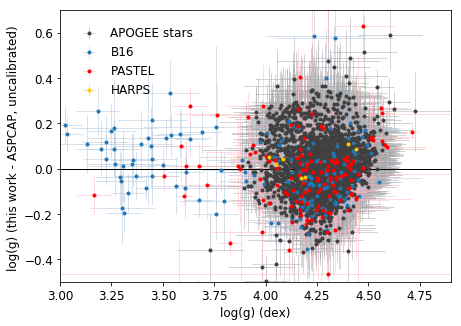}}
\caption{Diagram representing the uncorrected output $\log g$ from our pipeline (see Fig.\ref{HR_diagram}) and the difference between it and the values published by ASPCAP for each star in our sample. Errors displayed in both axis are our pipeline's estimates using the covariance matrix.}
\label{aspcap_logg}
\end{figure}

We display a comparison between published ASPCAP $\log g$ values and our raw output $\log g$ in Fig. \ref{aspcap_logg}. The plot shows that the distributions of both our $\log g$ values and the ASPCAP ones are quite similar, with a median difference of -0.007\,dex between them. The difference between both $\log g$ values is also around or below 0.4\,dex for most stars in the sample, showing that the values obtained with our method are actually consistent with the available ASPCAP ones. However, several studies comparing asteroseismic and spectroscopic $\log g$ have demonstrated that later values need to be calibrated \citep[e.g ][]{holtzman2015abundances,mortier2014correcting}. Also, the comparison with stellar isochrones points to the inaccuracy of spectroscopic $\log g$ values. Therefore, we decided to apply the correction as shown in Fig. \ref{HR_diagram} and explained in section \ref{FGK_results}.

\subsection{\label{litcomp}Other Literature comparisons}

\begin{figure}
\resizebox{\hsize}{!}{\includegraphics{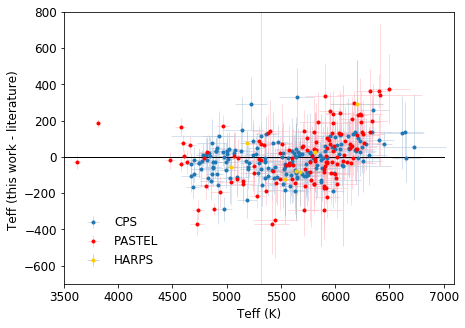}}
\caption{Diagram plotting the output $T_\mathrm{eff}$ of our pipeline against the difference between it and the values published by CPS, PASTEL, and the HARPS-GTO program. 
Errors included are our pipeline's estimates using the covariance matrix.}
\label{teff}
\end{figure}

\begin{figure}
\resizebox{\hsize}{!}{\includegraphics{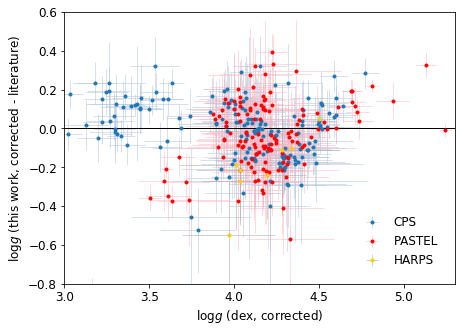}}
\caption{Diagram plotting the output $\log g_{cor}$ of our pipeline against the difference between it and the values published by CPS, PASTEL, and the HARPS-GTO program. 
Errors included are our pipeline's estimates using the covariance matrix.}
\label{logg}
\end{figure}

\begin{figure}
\resizebox{\hsize}{!}{\includegraphics{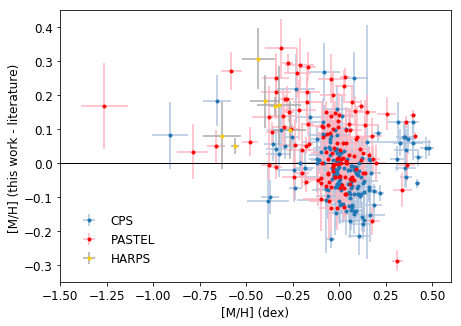}}
\caption{Diagram plotting the output $[M/H]$ of our pipeline against the difference between it and the values published by CPS, PASTEL, and the HARPS-GTO program. 
Errors included are our pipeline's estimates using the covariance matrix.}
\label{mh}
\end{figure}

\begin{table}
\caption{Average differences in derived parameters between our method and CPS parameters. Sample size is 168 stars. Median is shown to decrease effect of outliers. $\log g_{cor}$ value compared is our calibrated pipeline's output.}
\label{table:b16}
\centering
\begin{tabular}{c c}
\hline
Parameter & Median (Pipeline - CPS) \\ 
\hline
$T_\mathrm{eff}$ & -7\,K \\
$\log g_{cor}$ & +0.01\,dex\\
$[M/H]$ & -0.02\,dex \\
\hline
\end{tabular}
\end{table}

\begin{table}
\caption{Average differences in derived parameters between our method and PASTEL parameters. Sample size is 153 stars. Median is shown to decrease effect of outliers. $\log g_{cor}$ value compared is our calibrated pipeline's output.}
\label{table:pastel}
\centering
\begin{tabular}{c c c}
\hline
Parameter & Median (Pipeline - PASTEL) \\ 
\hline
$T_\mathrm{eff}$ & +7\,K \\
$\log g_{cor}$ & -0.06\,dex \\
$[M/H]$ & +0.04\,dex \\
\hline
\end{tabular}
\end{table}

The derived parameters for the stars in common with optical surveys are plotted in Figs. \ref{teff}-\ref{mh}, and are listed in Table \ref{table:CPSresults}.
The calculated differences between our method's parameters and the ones published in CPS, PASTEL, and HARPS-GTO are presented in Tables \ref{table:b16}- \ref{table:harps}.
In HARPS-GTO's case, and since there are only 8 stars in the sample, each individual result is included.

The spectra displayed in Fig. \ref{PASTEL_Gstar} show that our pipeline can minimize $\chi^2$ and match the APOGEE observed spectra with a synthesized spectrum to the required precision. Fig. \ref{teff} shows our derived $T_\mathrm{eff}$ for each of the three sub-samples of stars, plotted against the difference between it and literature parameters for the same stars. While the majority of the results have differences smaller than $200\,$K,
an upward trend is noticeable for stars above $6000\,$K, resulting in overestimation of temperatures for the hottest stars in the sample. The following Fig. \ref{logg} shows our derived $\log g_{cor}$ for each of the three sub-samples of stars, plotted against the difference between it and literature parameters for the same stars. This plot shows that, apart from our small subsample of subgiant stars, the surface gravity of most of our sample stars falls within a small range of values ($4.0$ to $4.6\,$dex) and $\Delta \log g$ ranging from -0.4 to +0.4 for these stars. 49 stars from CPS sample and 31 from PASTEL sample fall outside that range. For those stars, the $\Delta \log g$ values between our parameters and their literature values are different between samples, with our method resulting in values above literature for CPS stars and below for PASTEL. This difference can be explained by the usage of different methods for parameter determination. Fig. \ref{mh} shows our derived $[M/H]$ for each of the three sub-samples of stars, plotted against the difference between it and literature parameters for the same stars. A small gap appears, with our method reporting only 5 stars with $[M/H]$ between $+0.2$ and $+0.3$\,dex. This could be a statistical anomaly and due only to the number of stars in the sample. From the Figures \ref{teff}-\ref{mh} and the Tables \ref{table:b16} and \ref{table:pastel}, the method does not seem to have systematic errors in any of the three synthesized parameters ($T_\mathrm{eff}$, $\log g$, and $[M/H]$), as the error margins for our parameter estimates are within the optical measurements.

In the case of the PASTEL stars, some of the discrepancies can be explained by the fact that PASTEL presents $[Fe/H]$ and not $[M/H]$, and that their values for the $T_\mathrm{eff}$ come from different sources. This is not true for the CPS parameters, as they are uniformly calculated and have better precision than PASTEL parameters.

The largest difference between our derived parameters and literature is found for the $[M/H]$ measurements in the HARPS-GTO sample (average +0.16\,dex). This difference may be caused by either the sample size, as neither the larger PASTEL and CPS samples show these differences, or just a systematic error caused by the method used, as the HARPS-GTO parameters were measured using the EW method. The fact that HARPS-GTO results are $[Fe/H]$ and not $[M/H]$ may also have contributed to this difference. All the HARPS-GTO stars in common with our sample are relatively metal-poor, having $[M/H]$ ranging from -0.82 to -0.36\,dex, which means they are in a relatively small parameter space and may not represent a real distribution of stars like the other studied samples.

Our parameter estimates resulted in a high number of stars within the PASTEL sample for which the difference with respect to the PASTEL values was significant, with 21 out of 153 stars having parameters that differ significantly from literature values ($\left | \Delta [M/H] \right | > 0.35$\,dex; $\left | \Delta \log g \right | > 0.5$\,dex; $\left | \Delta T_\mathrm{eff} \right | > 400$\,K). The worst of these outliers are either very metal-poor stars ($[M/H]<-1.5\,$dex) or cold K and M dwarfs ($T_\mathrm{eff}<4000\,$K). Despite our pipeline not being optimized for these stars, their existence must be taken into account when synthesizing large samples of stars.

An additional explanation for any discrepancies can be the fact that APOGEE spectra is in the H-band (infrared) and CPS, PASTEL and HARPS-GTO results were calculated using spectra in the optical wavelength range.

\begin{table*}
\caption{Differences in derived parameters between our method and HARPS-GTO (our pipeline-literature). The H denotes the literature value for the parameter. $\log g_{cor}$ value compared is our calibrated pipeline's output.}
\label{table:harps}
\centering
\begin{tabular}{c c c c c c c c}
\hline
ID (2mass) & $ T_\mathrm{eff}$ (K) & $T_\mathrm{eff}^{H}$ & $\log g_{cor}$ (dex) & $\log g^{H}$ (dex) & $[M/H]$ (dex) & $[M/H]^{H}$ (dex)\\ 
\hline
\object{2M03402202-0313005} & 5762 & 5884 & 3.96 & 4.52 & -0.52 & -0.82 \\
\object{2M04042029-0439185} & 5192 & 5116 & 4.50 & 4.45 & -0.34 & -0.51 \\
\object{2M07385132-0527558} & 5635 & 5716 & 4.01 & 4.20 & -0.32 & -0.49 \\
\object{2M08172935-0359221} & 5678 & 5762 & 4.04 & 4.31 & -0.40 & -0.58 \\
\object{2M15074648+0852472} & 5811 & 5782 & 4.04 & 4.25 & -0.43 & -0.74\\
\object{2M15124763-0109577} & 5043 & 5096 & 4.34 & 4.44 & -0.56 & -0.61\\
\object{2M16302844+0410411} & 6200 & 5908 & 4.28 & 4.39 & -0.63 & -0.71\\
\object{2M16410822-0251258} & 5542 & 5665 & 4.19 & 4.43 & -0.26 & -0.36\\
\hline
\end{tabular}
\end{table*}

\subsection{\label{Fdis} Further Discussion}

As described in Section \ref{litcomp} and \ref{ASPCAP}, and Tables \ref{table:aspcap}-\ref{table:b16} and Figures \ref{aspcap_mh}-\ref{teff}, comparisons between previous observations in the optical and in the NIR for our sample stars show that there are differences between our parameters and the ones in literature. This section proposes and analyzes some explanations for these differences.

Errors in the line-list and/or the models used for the synthesis might be one reason for these discrepancies. The synthetic spectra match the observed one with a very high degree of accuracy (see figures \ref{solarspectra} and \ref{PASTEL_Gstar}), so it is unlikely this is the case. Tests with solar spectra show that our pipeline can provide accurate parameters for a solar-type star with APOGEE resolution (see section \ref{solar}), so it is unlikely that this is the explanation for all discrepancies found between this method's parameters and the ones previously derived in the literature.

Another additional hypothesis can be the difference between $[M/H]$ and $[Fe/H]$ (iron abundances) for these stars, as the CPS value we are comparing is $[M/H]$ and both PASTEL and HARPS-GTO samples published $[Fe/H]$ instead. \cite{valenti2005spectroscopic} found different distributions for these two parameters in nearby FGK stars, with $[M/H]$ having an average of 0.0\,dex and $[Fe/H]$ of +0.04\,dex. It should be noted, however, that \cite{haywood2001revision} has found $[Fe/H]=0$ for stars in the solar neighborhood, and \cite{rocha1996metallicity} measured an average of $[Fe/H]=-0.2$\,dex for G dwarfs in the solar neighborhood. Therefore, comparing the overall metallicities for each star with their iron abundances can lead to unwanted disparities. We determined $[Fe/H]$ for each star in the sample using our pipeline, and compared them to values in literature. These syntheses were done having only the Fe abundance as a free parameter and fixing the other parameters to the results previously derived by our pipeline. The abundances derived are summarized in Fig. \ref{FeH} and are available in Table \ref{table:CPSresults} in the appendix.

\begin{figure}
\resizebox{\hsize}{!}{\includegraphics{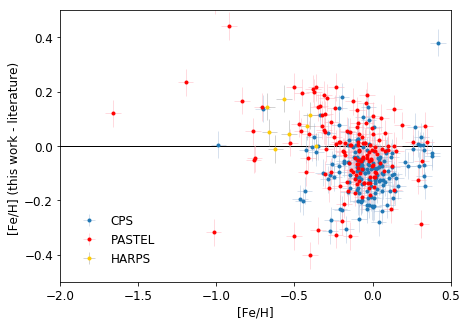}}
\caption{Comparison between iron abundance derived by our pipeline and literature values.}
\label{FeH}
\end{figure}

\begin{table}
\caption{Average differences between $[Fe/H]$ derived with this method and literature values for $[Fe/H]$. Median is shown to decrease effect of outliers.}
\label{table:feh}
\centering
\begin{tabular}{c c c c}
\hline
Sample & Stars & Median (Pipeline - Literature)\\ 
\hline
CPS & 168 & -0.08\,dex \\
PASTEL & 153 & -0.02\,dex \\
HARPS-GTO & 8 & +0.06\,dex\\
\hline
\end{tabular}
\end{table}

In Table \ref{table:feh}, it is still clear that the $[Fe/H]$ value given by our pipeline is very close to the literature values. The largest measured differences are found when comparing our values with CPS ($\Delta [Fe/H] = -0.08\,$dex), while the median difference in the PASTEL sample comparison is actually -0.02\,dex. In the case of HARPS-GTO stars, our results are now in better agreement with theirs ($\Delta [Fe/H] = +0.06\,$dex), as we are comparing $[Fe/H]$ rather than $[M/H]$. These results show us that the method is robust and can provide not only $[M/H]$ but $[Fe/H]$ as well for the analyzed stars.

\begin{figure}
\resizebox{\hsize}{!}{\includegraphics{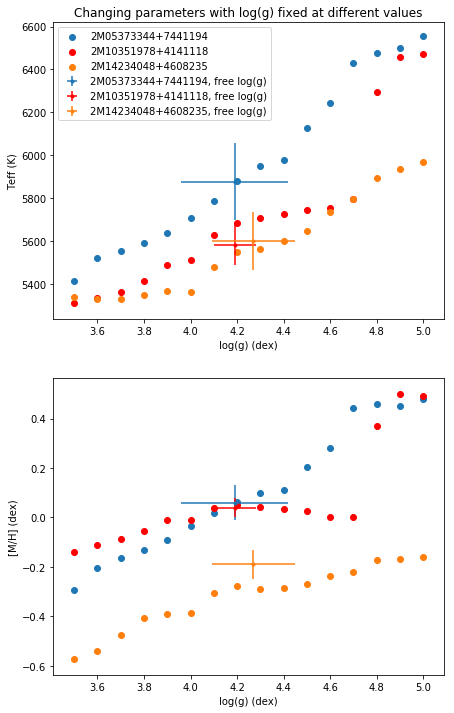}}
\caption{Diagram representing the output parameters with fixed surface gravity at different values between $\log g=3.5\,$dex and $\log g=5.0\,$dex. The data for 3 different stars is presented, 2M05373344+7441194 (blue), 2M14234048+4608235 (orange), and 2M10351978+4141118 (red), with the measured values for each of them indicated by an errorbar of the same color.}
\label{fixed_logg}
\end{figure}

The fact that the spectroscopic $\log g$ need to be revised make us wonder what is the impact of an incorrect $\log g$ on the simultaneous derivation of $T_\mathrm{eff}$ and $[M/H]$. Syntheses with lower values for $\log g$ can result in lower values as well for $T_\mathrm{eff}$ and $[M/H]$ across a sample of stars, as the line broadening due to low surface gravity must be compensated by a corresponding decrease in $[M/H]$. This effect is in turn compensated by a decrease in $T_\mathrm{eff}$ that results in wider lines. ASPCAP's calibrations correct some of these effects, increasing $T_\mathrm{eff}$ by around $+90$\,K and $[M/H]$ by $+0.027$\,dex for stars with $[M/H]>-0.5$\,dex. The differences found when comparing our output $T_\mathrm{eff}$ and $[M/H]$ with the ASPCAP values might be explained by the fact that, unlike their pipeline, our values for these parameters are not calibrated in any way.

Regarding the surface gravity differences between our work and ASPCAP results, we decided to run new syntheses with our method for three test stars, representatives of the full sample. We fixed surface gravity to different values between $\log g=3.5\,$dex and $\log g=5.0\,$dex, and compared the resulting changes in the other spectroscopic parameters. Since both our pipeline and ASPCAP's use similar methods for spectroscopic parameter derivation, by evaluating the magnitude of the differences, we can conclude how problems in the derivation of $\log g$ by both our method and ASPCAP affect other parameters. The results obtained for the 3 test stars are presented in Fig. \ref{fixed_logg}. Comparing our test to a similar one with fixed $\log g$ values published in \cite{tsantaki2018atmospheres}, we find that our results are similar to theirs, although with a larger dispersion in both $T_\mathrm{eff}$ and $[M/H]$. The derived $T_\mathrm{eff}$ can vary by up to 1000\,K, with a strong dependence on $\log g$. There is also a strong dependence on the output for $[M/H]$ as $\log g$ is fixed at different values, with differences varying up to 0.75\,dex. The larger magnitude of these differences can be explained by the lower resolution of the analyzed spectra here as compared to the spectra used by \cite{tsantaki2018atmospheres} to test the effects of fixing the $\log g$. We note that this interdependence of parameters reported for the spectral synthesis method is not relevant when using the ionization and excitation balance method (see also \cite{mortier2013new} and \cite{mortier2014correct} for a further exploration of the effects fixing $\log g$ values can have on other spectroscopical parameters).

This test shows that performing syntheses with wrong values for the surface gravity can result in errors across other parameters, and that fixing the $\log g$ can result in different parameters for the same star. Despite the strong effect changing the $\log g$ can have on the other parameters, such as $T_\mathrm{eff}$ and $[M/H]$, the fact that our uncorrected $\log g$ are not very different from ASPCAP and optical values, and that our $T_\mathrm{eff}$ and $[M/H]$ are also comparable, increases our confidence in our final set of parameters.

Another possible explanation for any discrepancies in the parameters may be just a difference in the way the parameters are minimized in our pipeline and other codes. For example, the work of \cite{jofre2014gaia} explored the derivation of stellar parameters of the same stellar sample by multiple codes and demonstrated how the metallicities derived by \textit{Turbospectrum} can be lower than values obtained with other methods. In addition, the recent work of \cite{ispec2019sbc} has shown how different spectral synthesis codes can derive different parameters with iSpec. This work focuses on optical wavelengths, but some differences, such as the ones described in Section 3.8 of that paper, ``The One variable at a time experiment'', are independent of the wavelength used. Further explorations of the output parameter space of \textit{Turbospectrum} and other codes using NIR spectra are needed to know exactly what the explanation for these discrepancies might be.

Finally, even though our synthetic spectra provide a good match for the observed spectra, it does not mean we can fully trust the parameters or that they are better or worse than the ones published before in the literature. Since the observations are different from the ones in the optical, and the normalization is different from the one used by ASPCAP as well, there is no way to conclusively determine the set of parameters that better characterize a particular star. We can state for certain that all the parameters shown in this paper were derived in a homogeneous way, using a uniform method and pipeline, and obtaining precise results. Therefore, any systematic error or bias will be present across the full sample.

\section{\label{conclusions} Conclusions}

Using synthetic spectra, we provide parameters for a sample of 3748 FGK main-sequence and subgiant stars observed with APOGEE. These parameters are derived using synthetic spectra generated by a pipeline using both \textit{iSpec} and \textit{Turbospectrum}. This method requires a functional and complete line-list and complementing line mask for the H-band, which are included for future reference. The results are compared with multiple literature sources, and possible sources of discrepancies are explored.

Repeated iterations of \textit{Turbospectrum} syntheses with similar spectra show how the method can provide consistent output parameters across multiple runs. The matching spectral synthesis and precise parameter determinations show that our pipeline can be a very powerful tool to synthesize stellar spectra and derive spectroscopic parameters for stars with H-band spectra. The fact that our synthesized spectra are a very close match to the observations (see figure \ref{PASTEL_Gstar}) and that the average differences in derived parameters between our measurements and others are low (and below precision level in the specific case of comparison with optical values),
means that this method can be a viable alternative in the future for homogeneous spectroscopic parameter derivation in the H-band, and may be expanded to other spectral types and spectral resolutions.

Even though this paper's focus is on FGK dwarfs, the most promising next step is to move towards colder stars and derive parameters for M dwarfs. Another paper exploring stars in this parameter space is currently in preparation.

\begin{acknowledgements}
 This work was supported by FCT - Fundação para a Ciência e a Tecnologia through national funds (PTDC/FIS-AST/28953/2017, 
 PTDC/FIS-AST/7073/2014, PTDC/FIS-AST/32113/2017, UID/FIS/04434/2013) 
 and by FEDER - Fundo Europeu de Desenvolvimento Regional through COMPETE2020 - 
 Programa Operacional Competitividade e Internacionalização 
 (POCI-01-0145-FEDER-028953, POCI-01-0145-FEDER-016880, POCI-01-0145-FEDER-032113, POCI-01-0145-FEDER-007672).
 This work was supported by FCT/MCTES through national funds (PIDDAC) by grant UID/FIS/04434/2019.
 This research has made use of NASA’s Astrophysics Data System.
 P.S. acknowledges the support by the Bolsa de Investigação PD/BD/128050/2016.
 E.D.M. acknowledges the support by the Investigador FCT contract IF/00849/2015/CP1273/CT0003 and in the form of an exploratory project with the same reference.
 B.R-A acknowledges funding support from CONICYT PAI/CONCURSO NACIONAL INSERCION EN LA ACADEMIA CONVOCATORIA 2015 grant 79150050 and FONDECYT through grant 11181295.
\end{acknowledgements}

\bibliographystyle{aa}
\bibliography{/home/psarmento/Dropbox/Doutoramento/Doutoramento/Paper1_Sep2018/notes.bib}

\begin{multicols}{2}
\end{multicols}
\begin{appendix}

\section{Air/vacuum wavelength}

APOGEE publishes its spectra in vacuum wavelength \citep{albareti2016}, while the spectral lines in our used line-list are all in air wavelength. 
In order to use the lines in conjunction with APOGEE's spectra, they must be converted into the same format.
Thus, a good conversion in the H-band spectra is essential to use iSpec's capabilities with APOGEE's spectra. 
This conversion is done using equations found in ``Conversion from vacuum to standard air wavelengths'' by \cite{prieto2011}. 
The equation used was:
\begin{equation}
 \frac{\lambda _{0}-\lambda}{\lambda }=n-1=a+\frac{b_{1}}{c_{1}-1/\lambda _{0}^{2}}+\frac{b_{2}}{c_{2}-1/\lambda _{0}^{2}}
\end{equation}
, where $\lambda _{0}$ is given in $\mu$m and the parameters used were the ones in Table \ref{table:2} \citep{ciddor1996refractive}.

\begin{table}
\caption{Parameters used for air-vacuum conversion}
\label{table:2}
\centering
\begin{tabular}{c c}
\hline
  Parameters & Value \\ 
\hline
a & 0.0 \\
$b_{1}$ & $5.792105\times 10^{-2}$ \\
$b_{2}$ & $1.67917\times 10^{-3}$ \\
$c_{1}$ & 238.0185 \\
$c_{2}$ & 57.362 \\
\hline
\end{tabular}
\end{table}

\section{Additional synthesized spectra examples}

\begin{landscape}

\begin{figure}
\resizebox{\hsize}{!}{\includegraphics{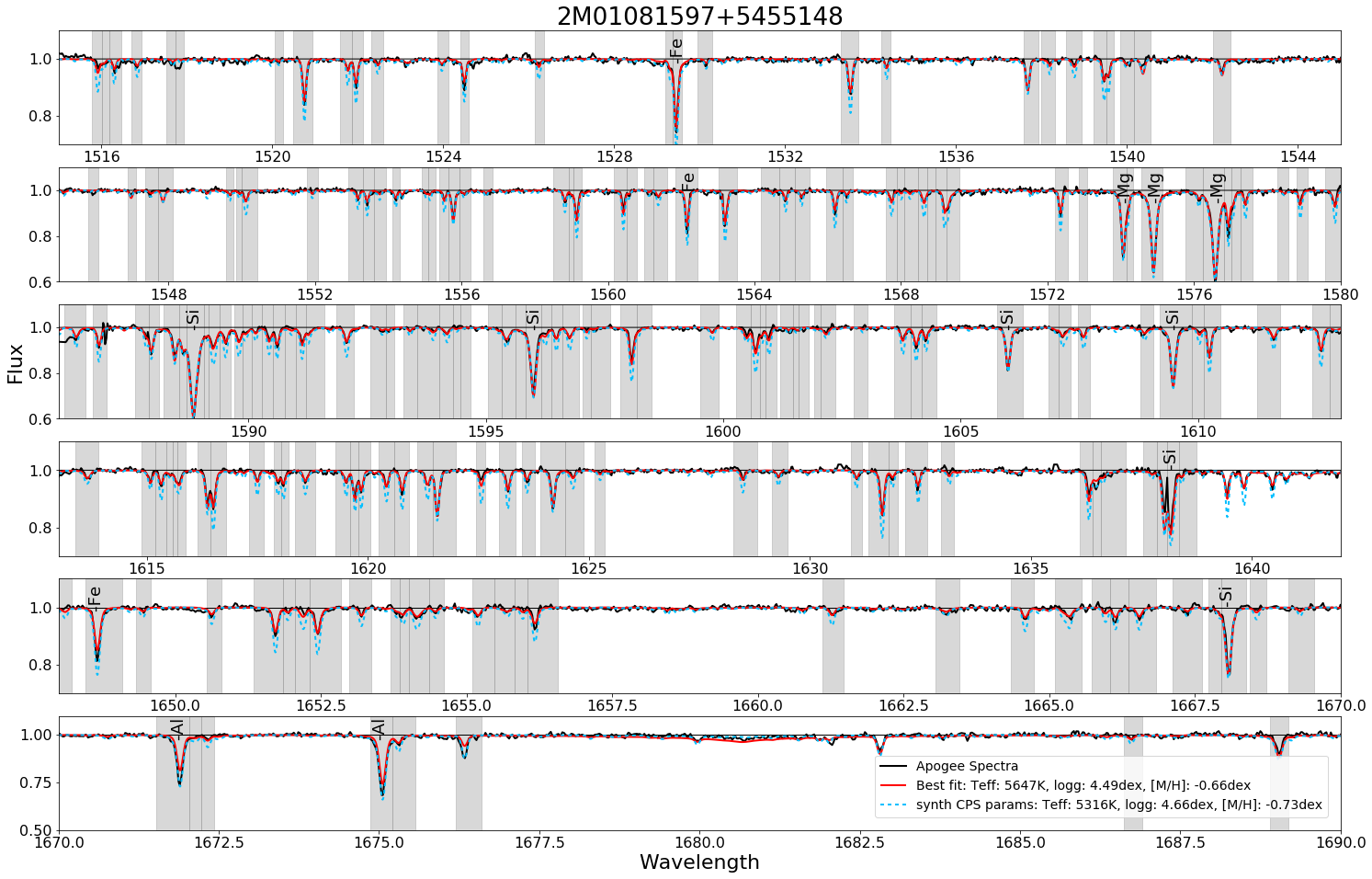}}
\caption{Comparison between APOGEE spectra of the star 2M01081597+5455148 (HD6582, black) and two synthetic spectra (red, straight line,
with the our pipeline's best match parameters and blue, dashed, with the CPS parameters for this star) for APOGEE wavelength range. 
In grey highlight are the areas used for $\chi ^2 $ minimization by our pipeline's algorithm.
The best match parameters derived were $T_\mathrm{eff}=5639\pm163$\,K,$\log g=4.46\pm0.20$\,dex, $[M/H] = -0.66\pm0.07$\,dex, 
and the ones published in CPS were $T_\mathrm{eff}=5316\pm25$\,K,$\log g=4.66\pm0.028$\,dex, $[M/H] = -0.73\pm0.01$\,dex.}
\label{cold_spectrum}
\end{figure}

\end{landscape}

\begin{landscape}

\begin{figure}
\resizebox{\hsize}{!}{\includegraphics{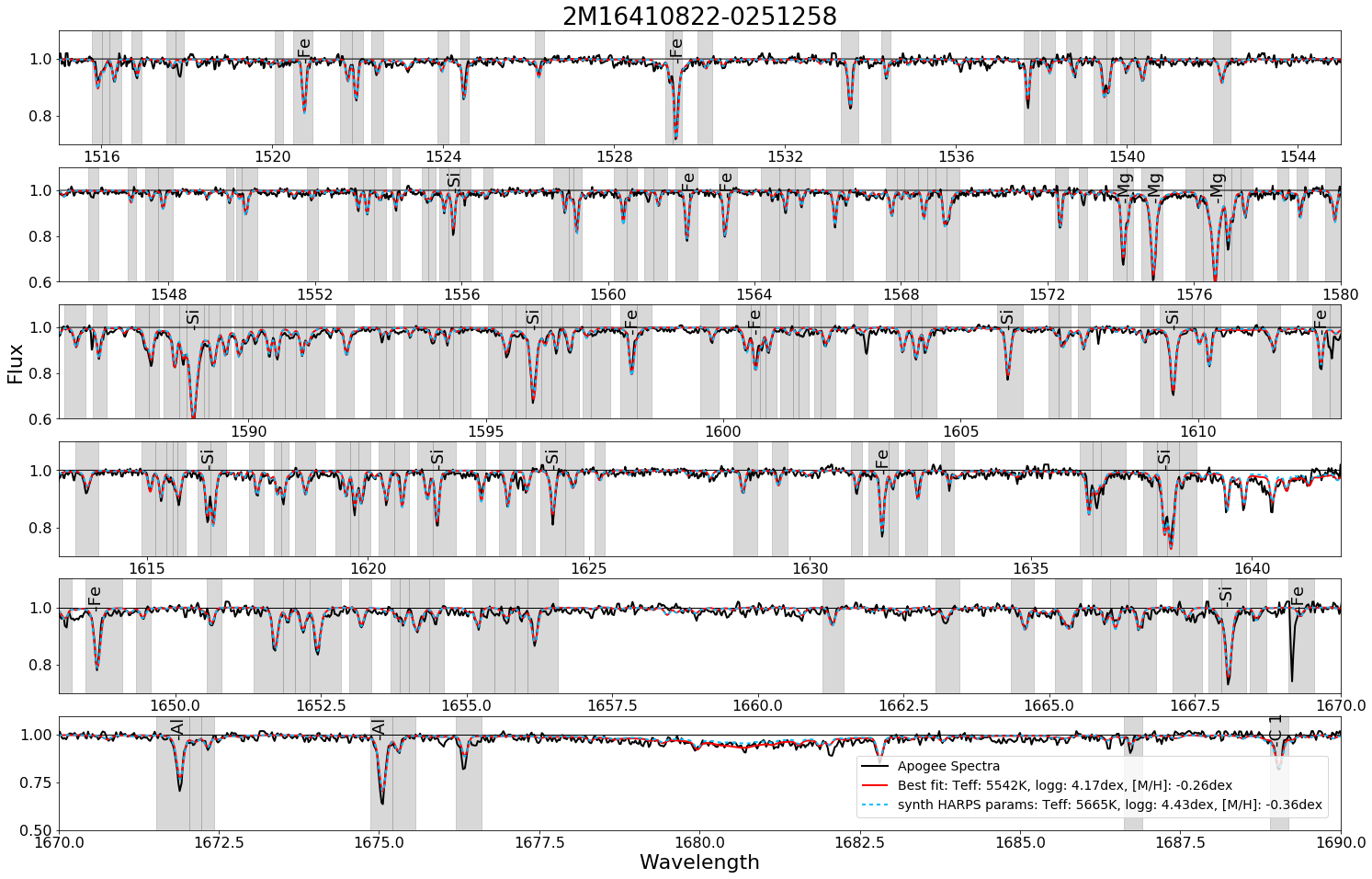}}
\caption{Comparison between APOGEE spectra of the star 2M16410822-0251258 (HD 150433, black) and two synthetic spectra (red, straight line, 
with the our pipeline's best match parameters and blue, dashed, with the HARPS-GTO parameters for this star) for APOGEE wavelength range. 
In grey highlight are the areas used for $\chi ^2 $ minimization by our pipeline's algorithm.
The best match parameters derived were $T_\mathrm{eff}=5547\pm201$\,K,$\log g=4.17\pm0.23$\,dex, $[M/H] = -0.26\pm0.09$\,dex, 
and the ones published by the HARPS-GTO were $T_\mathrm{eff}=5665\pm12$\,K,$\log g=4.43\pm0.02$\,dex, $[M/H] = -0.36\pm0.01$\,dex.}
\label{high_spectrum}
\end{figure}

\end{landscape}

\begin{landscape}

\begin{figure}
\resizebox{\hsize}{!}{\includegraphics{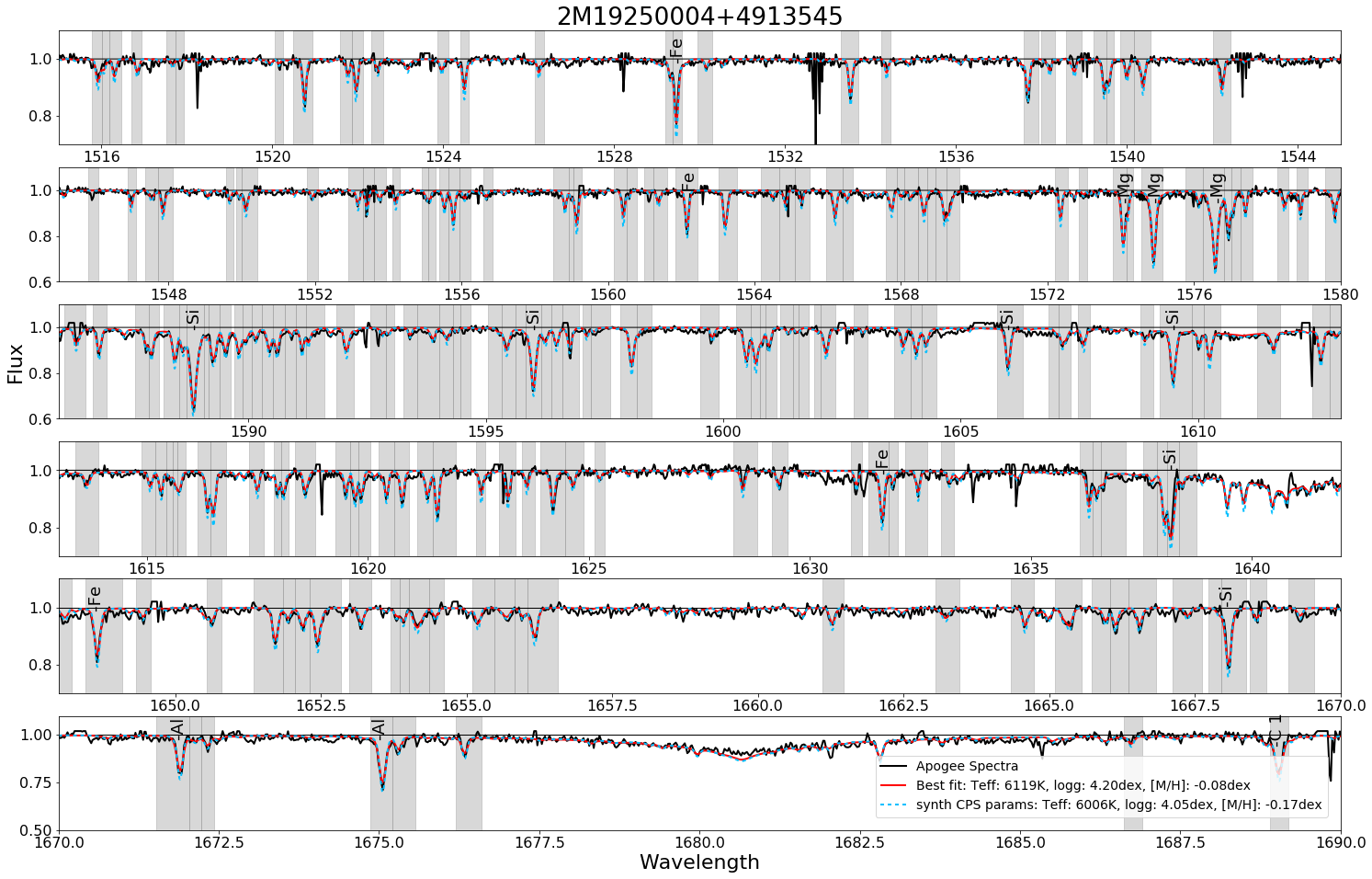}}
\caption{Comparison between APOGEE spectra of the star 2M19250004+4913545 (Kepler-36, black) and two synthetic spectra (red, straight line, 
with the our pipeline's best match parameters and blue, dashed, with the CPS parameters for this star) for APOGEE wavelength range. 
In grey highlight are the areas used for $\chi ^2 $ minimization by our pipeline's algorithm.
The best match parameters derived were $T_\mathrm{eff}=6127\pm213$\,K,$\log g=4.21\pm0.31$\,dex, $[M/H] = -0.07\pm0.09$\,dex, 
and the ones published in CPS were $T_\mathrm{eff}=6006\pm25$\,K,$\log g=4.05\pm0.028$\,dex, $[M/H] = -0.17\pm0.01$\,dex.}
\label{hot_spectrum}
\end{figure}

\end{landscape}

\onecolumn
\begin{landscape}

\section{Full results for stellar sample}

\begin{longtable}{cccccccccc}
 \caption{\label{table:CPSresults}Excerpt of the list of derived parameters for FGK stellar sample. Uncertainties in derived parameters are estimated by iSpec. 
\textit{iSpec} output parameters indicated by letter O. Literature parameters indicated by letter L. Full list available for download.}\\
 \hline\hline
APOGEE\_ID & $T_\mathrm{eff} \pm \Delta T_\mathrm{eff}$ &
 $\log g \pm \Delta \log g$ & $\log g_{cor}$ & $\textrm{[M/H]} \pm \Delta \textrm{[M/H]}$ & $[Fe/H] \pm \Delta \textrm{[Fe/H]}$ &
 v$_\mathrm{mic}$ & v$_\mathrm{mac}$ &  $v \sin i$  &$\chi ^2$ \\
\hline
\endfirsthead
\caption{continued.}\\
\hline\hline
APOGEE\_ID & $T_\mathrm{eff} \pm \Delta T_\mathrm{eff}$ &
 $\log g \pm \Delta \log g$ & $\log g_{cor}$ & $\textrm{[M/H]} \pm \Delta \textrm{[M/H]}$ & $[Fe/H] \pm \Delta \textrm{[Fe/H]}$ &
 v$_\mathrm{mic}$ & v$_\mathrm{mac}$ &  $v \sin i$  &$\chi ^2$ \\
\hline
\endhead
\hline
\endfoot
\object{2M00012723+8520108}	& $	5973	\pm	66	$ & $	4.24	\pm	0.07	$ & $	4.15	$ & $	0.16	\pm	0.03	$ & $	0.11	\pm	0.1	$ &	0.74	&	5	&	10.11	&	1.4	\\
\object{2M00015324+5634361}	& $	6164	\pm	143	$ & $	4.15	\pm	0.2	$ & $	4.01	$ & $	-0.15	\pm	0.06	$ & $	-0.23	\pm	0.1	$ &	0.64	&	6.54	&	8.57	&	0.54	\\
\object{2M00072254+2627025}	& $	4741	\pm	83	$ & $	3.18	\pm	0.11	$ & $	3.18	$ & $	-0.03	\pm	0.04	$ & $	-0.15	\pm	0.1	$ &	0.88	&	3.67	&	9.97	&	1.02	\\
\object{2M00095611-0002296}	& $	6196	\pm	139	$ & $	4.47	\pm	0.15	$ & $	4.31	$ & $	0.02	\pm	0.05	$ & $	-0.05	\pm	0.1	$ &	0.21	&	6.54	&	8.48	&	1.41	\\
\object{2M00100176+0201021}	& $	6142	\pm	115	$ & $	4.26	\pm	0.16	$ & $	4.12	$ & $	0.09	\pm	0.04	$ & $	0.02	\pm	0.1	$ &	0.67	&	6.11	&	8.99	&	1.46	\\
\object{2M00125570-1441121}	& $	5730	\pm	111	$ & $	4.3	\pm	0.12	$ & $	4.27	$ & $	0.04	\pm	0.05	$ & $	0	\pm	0.1	$ &	0.65	&	4	&	9.3	&	0.44	\\
\object{2M00135646-1439234}	& $	5927	\pm	127	$ & $	4.33	\pm	0.19	$ & $	4.25	$ & $	-0.15	\pm	0.05	$ & $	-0.21	\pm	0.1	$ &	0.16	&	5.08	&	8.3	&	1.46	\\
\object{2M00153822-1516023}	& $	6181	\pm	145	$ & $	4.52	\pm	0.15	$ & $	4.37	$ & $	0.09	\pm	0.06	$ & $	0.04	\pm	0.1	$ &	0.52	&	6.35	&	10.49	&	1.7	\\
\object{2M00162970-1456483}	& $	6144	\pm	122	$ & $	4.36	\pm	0.19	$ & $	4.22	$ & $	0.11	\pm	0.05	$ & $	0.06	\pm	0.1	$ &	0.68	&	6.08	&	10.61	&	0.48	\\
\object{2M00164203-1533376}	& $	5852	\pm	140	$ & $	4.23	\pm	0.2	$ & $	4.17	$ & $	-0.18	\pm	0.06	$ & $	-0.27	\pm	0.1	$ &	0.36	&	4.76	&	8.23	&	0.48	\\
\object{2M00170412-1459180}	& $	5598	\pm	106	$ & $	4.27	\pm	0.13	$ & $	4.27	$ & $	-0.01	\pm	0.05	$ & $	-0.04	\pm	0.1	$ &	0.88	&	3.62	&	9.98	&	0.56	\\\hline

\end{longtable}

\end{landscape}
\npnoround

\end{appendix}

\end{document}